\documentstyle[sprocl]{article}

\input{psfig}

\def\Journal#1#2#3#4{{#1} {\bf #2}, #3 (#4)}

\def\PRL{\em Phys. Rev. Lett.}
\def\PRD{{\em Phys. Rev.} D}

\def\be{\begin{equation}}
\def\ee{\end{equation}}
\def\bea{\begin{eqnarray}}
\def\eea{\end{eqnarray}}
\begin{document}

\title{FIRST-ORDER PHASE TRANSITIONS IN AN EARLY-UNIVERSE ENVIRONMENT}

\author{MATTHEW LILLEY}\address{DAMTP,
  Silver Street, Cambridge,\\ CB3 9EW, UK\\E-mail:
  M.J.Lilley@damtp.cam.ac.uk}


\maketitle\abstracts{In first-order phase transitions in the early
  universe, the bubble wall is expected to be significantly
  slowed-down by its interaction with the surrounding plasma.  We
  examine the behaviour of the phase of the Higgs field after
  two-bubble collisions, and find that phase differences equilibrate
  much more quickly in slow-moving bubbles than in those which expand
  at the speed of light.  This could lead to a significant reduction
  in the initial density of topological defects formed at a
  first-order phase transition.}

\section{Introduction}
First-order phase transitions proceed by bubble nucleation and
expansion.  When three or more bubbles collide a phase winding of
$2\pi n$ may be generated, forming a cosmic string in the region
between them.  In order to assess the cosmological significance of
cosmic strings, it is important to be able to forecast their initial
density.  This depends on the behaviour of the phase of the Higgs
field after bubbles collide and merge -- in particular if the phase
difference between two bubbles can equilibrate before the arrival of
the crucial third bubble, there may be a strong suppression of the
initial string density.

At any phase transition where particles acquire mass, those particles
outside the bubble without enough energy to become massive inside
bounce off of the bubble wall, retarding its progress through the
plasma.  The faster the bubble is moving, the greater the momentum
transfer in each collision, and hence the stronger the retarding
force.  Thus a force proportional to the bubble-wall velocity appears
in the effective equations of motion. Impeded by its interaction with
the hot plasma, the bubble wall reaches a terminal velocity $v < c$ --
for the (Standard Model) electroweak phase transition, the value
$v\sim 0.1c$ was predicted \cite{mp}. In this paper, we investigate
the consequences of slow-moving bubble walls on phase equilibration in
global- and local-symmetry models.

\section{Phase Equilibration}
Writing the Higgs field $\Phi=\rho e^{i\theta}$ the equations of
motion for the Abelian Higgs ($U(1)$ gauge symmetry) model (which we
consider, for simplicity) are
\begin{eqnarray}
\label{eq:local_rho} \ddot{\rho} - {\rho}'' - ({\partial}_{\mu}\theta -
  e{A}_{\mu})^{2}\rho & = &-\frac{\partial V}{\partial \rho}\\
\label{eq:local_theta} \partial^{\mu}\bigl[\rho^{2}(\partial_{\mu}\theta
- eA_{\mu})\bigr] &=& 0\\ 
\label{eq:local_a}\ddot{A_\nu} - {A_\nu}'' -
{\partial}_{\nu}\left(\partial
 \cdot A \right) & = & -2e \rho^2 {\partial}_{\nu}\theta.
\end{eqnarray}
Taking, after Kibble and Vilenkin \cite{kv}, our gauge-invariant phase
difference between two points to be
\begin{equation}\label{eq:gipd}
\Delta \theta = \int_{A}^{B} dx^{i}\left(\partial_{i} - i e A_{i}\right),
\end{equation}
it is possible to derive an analytic expression for the phase
difference after time $t$ between the centres of two bubbles nucleated
at time zero with radius $R$ and initial phase difference $2\theta_0$
\begin{equation} \label{eq:kvgipd}
\Delta \theta = \frac{2 R}{t} \theta_{0} \left( \cos e\eta\left(t - R\right)
    + \frac{1}{e \eta R}\sin e \eta\left(t - R\right)\right),
\end{equation}
that is, decaying {\em phase oscillations} take place -- see
\cite{cst} for numerical verification.

In order to model the interaction of the bubble wall with the plasma,
we add a term $\Gamma \dot{\rho}$ to the equation of motion for the
modulus of the Higgs field, Eq. (\ref{eq:local_rho}), as motivated in
\S 1.  If there are no gauge fields, this leads to a different kind of
decaying phase oscillations \cite{fm}.  What happens in theories with a
gauge symmetry, where the bubbles move with speeds less than that of
light?

Eq. (\ref{eq:kvgipd}) was obtained by imposing $SO(1,2)$ Lorentz
symmetry on the field equations for the two-bubble problem.  If the
bubbles do not move at the speed of light, no such assumption is
possible.  This is because whilst the modulus $\rho$ of the Higgs
field is constrained to propagate at a speed $v$, there is no such
restriction on the phase $\theta$ or the gauge fields.  The problem
must then be approached via numerical simulations.

\section{Results}
For the sake of clarity, we have chosen to present our results in
terms of the evolution with time of the gauge-invariant phase
difference $\Delta \theta$ between the centres of the two bubbles,
though the qualitative behaviour was found not to change when
calculated between different points.

\begin{figure}[t]
  \centerline{ \psfig{figure=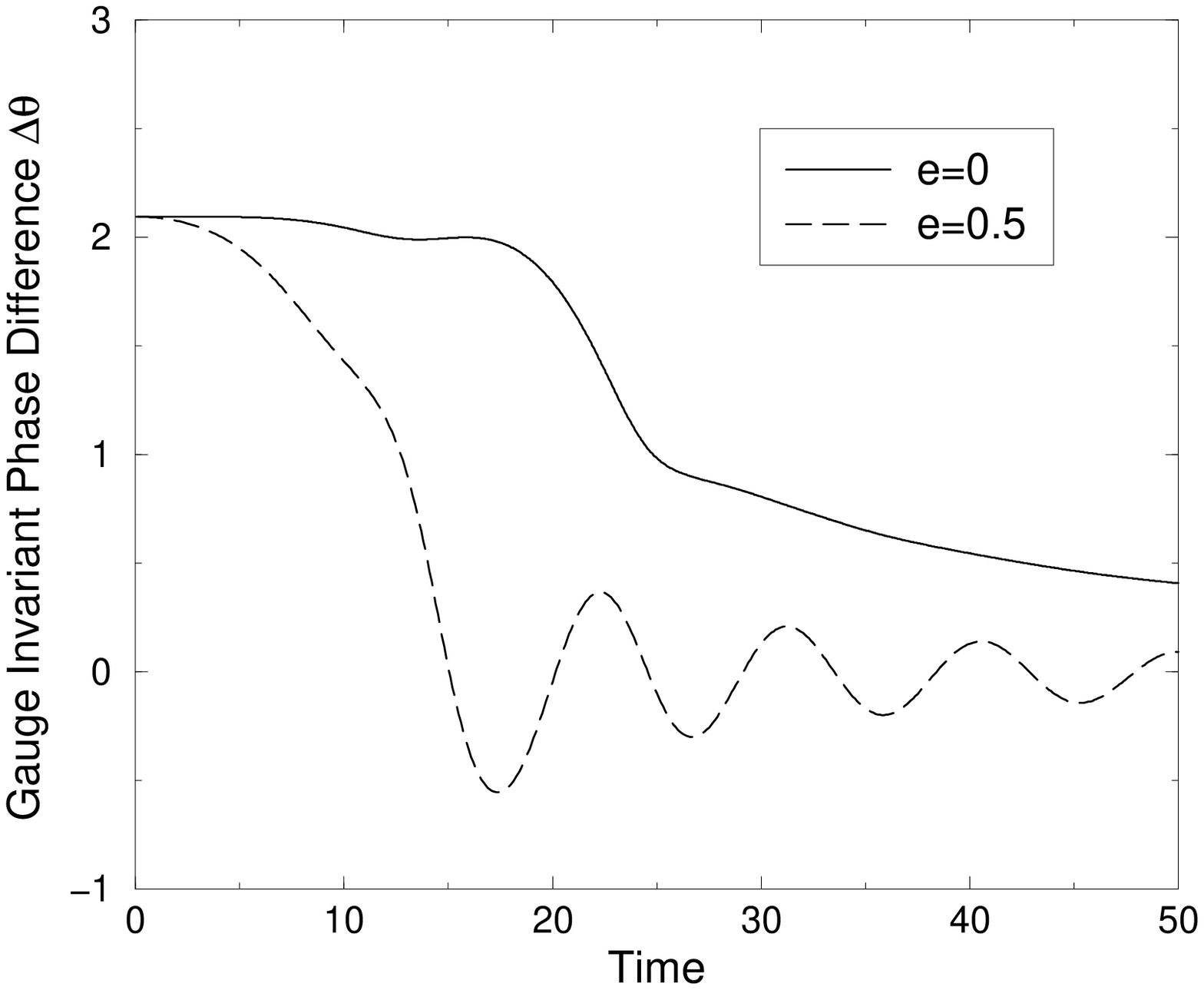,width=6 cm}
    \psfig{figure=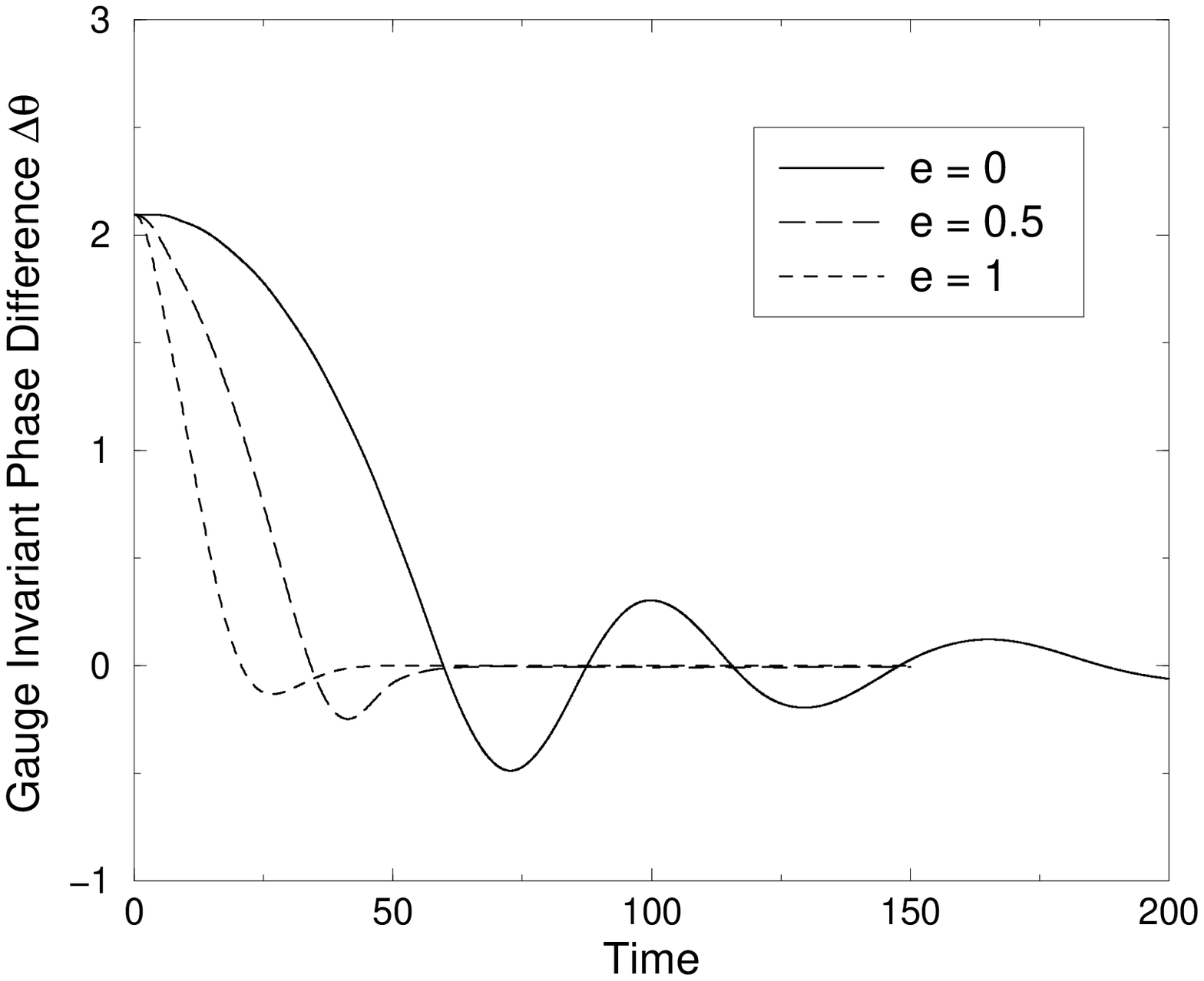,width=6 cm}}
\caption{Gauge-invariant phase difference between (a) two bubbles moving
  at the speed of light , $\Gamma = 0$, and (b) two slow-moving
  bubbles, $\Gamma = 2$.  \label{fig:phasegraph}}
\vskip 1 cm
  \centerline{\psfig{figure=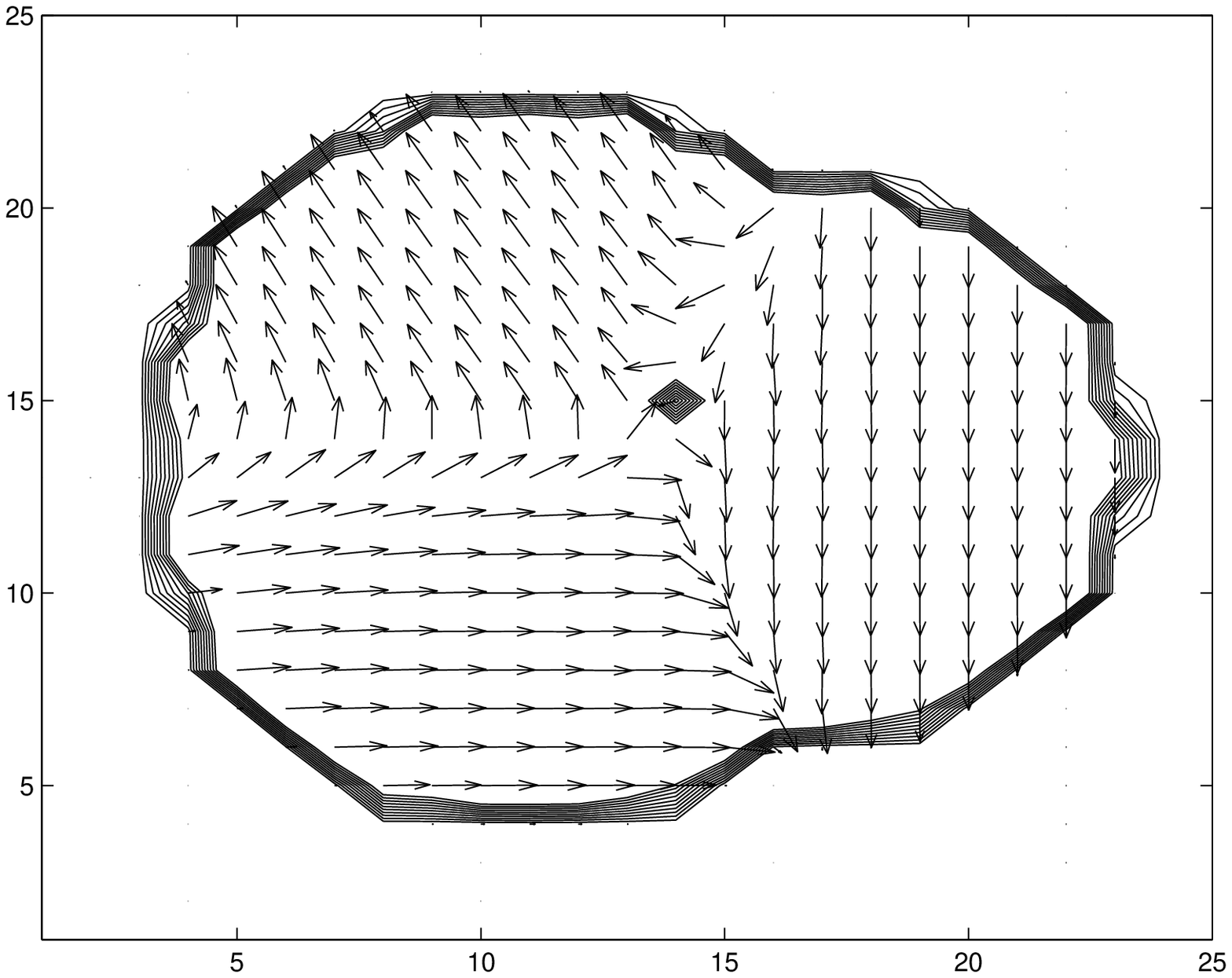,width=5.8 cm}
    \psfig{figure=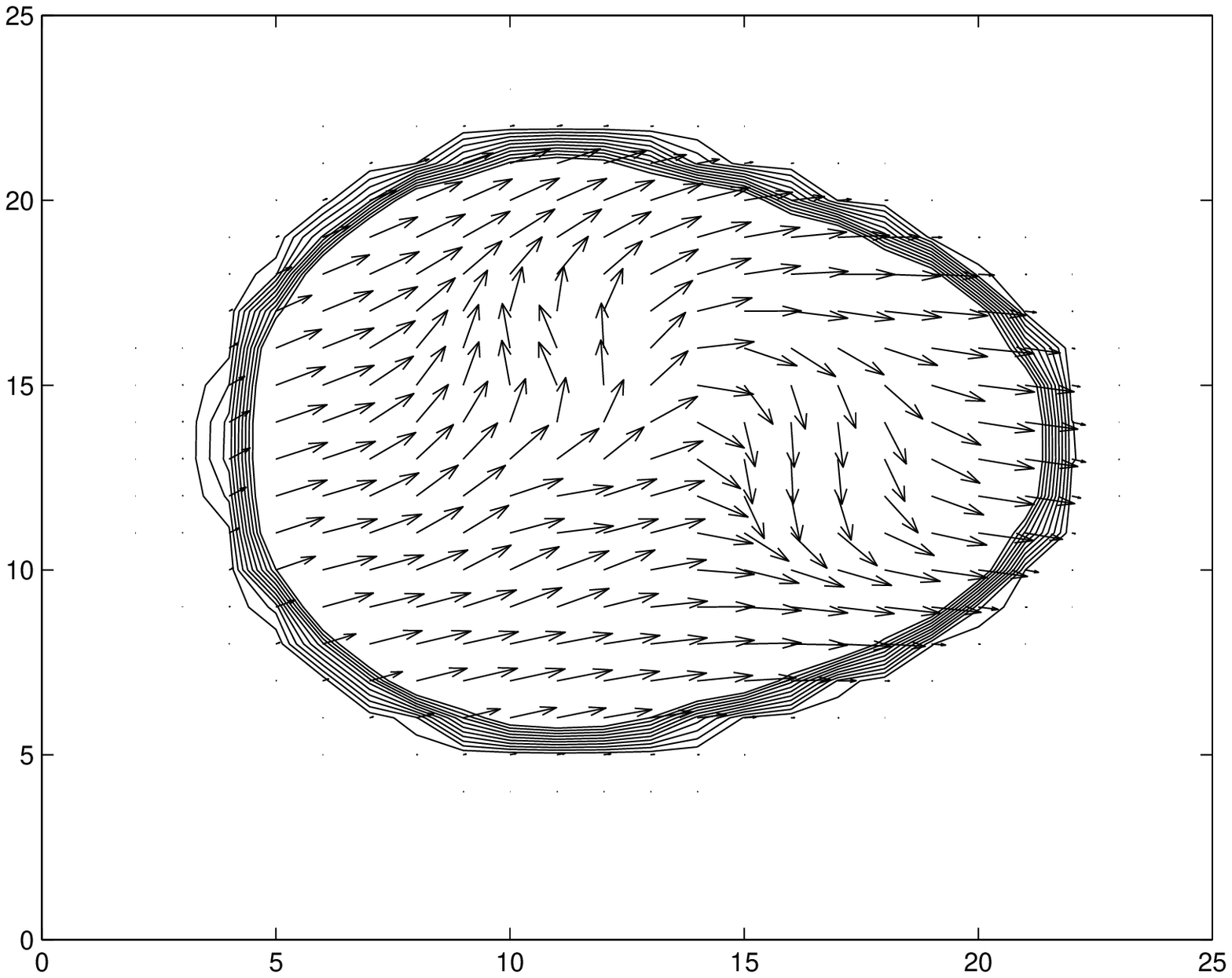,width=5.8 cm}}
\caption{Phase plot and bubble walls after three-bubble collisions,
  with phases 0 (bottom left), $2\pi / 3$ (top left) and $-\pi / 2$
  (right): (a) with $\Gamma = 0$ a vortex is formed at the centre, and
  (b) with identical initial conditions, but $\Gamma = 0.5$ there is no
  vortex. \label{fig:vortex}}
\end{figure}

Figure \ref{fig:phasegraph} (a) shows the behaviour of the
gauge-invariant phase difference for bubbles moving at the speed of
light -- the decaying oscillations calculated by Kibble and Vilenkin
in the local case.  In the global case, $e=0$, we find that the phase
{\em does} equilibrate, but on a much longer time-scale.  Thus we
would expect that for fast-moving bubbles, fewer defects are formed in
local theories than global ones, since in order to form a defect a
phase difference inside the two merged bubbles must still be present
when a third bubble collides.

In Figure \ref{fig:phasegraph} (b) we plot $\Delta \theta$ for
slower-moving bubbles.  For $e=0$, we confirm in $3+1$-dimensions the
decaying phase oscillations described by Ferrera and Melfo \cite{fm}
and observed by them in $2+1$-dimensions.  These oscillations are
killed by adding in gauge fields -- for a fixed bubble-wall velocity,
the stronger the gauge coupling, the less time the gauge-invariant
phase difference is non-zero, and hence the less likely a third
collision will occur in time for a defect to form.  Thus we would
expect a lower defect-formation rate in local theories with
slower-moving bubble walls.

Figure \ref{fig:vortex} illustrates our findings -- it shows a
cross-section through a non-simultaneous three-bubble collision, after
all three bubbles have merged.  In each case, the bubbles of initial
radius $R=5$, centred at $(\pm 8,0,-10)$ and $(0,0,10)$, were given
phases $\theta= {-\pi / 2}, {0}$ and ${2 \pi /3}$.  For identical
initial conditions, we see that in the fast-moving case a vortex is
formed, but when the bubbles are slowed down, the phase difference
between the two bubbles has equilibrated by the time the third bubble
collides, and no defect is formed.

For a fuller discussion of these and other results, including the
effect of taking into account the finite conductivity of the plasma,
and the magnetic fields formed at collisions of fast- and slow-moving
bubbles, see \cite{lilley}.

\section*{Acknowledgments}
This work was done in collaboration with A.-C. Davis.  We would like
to thank O.~T\"{o}rnkvist for helpful discussions.  Financial support
was provided by PPARC and Fitzwilliam College, Cambridge.  Computer
facilities were provided by the UK National Cosmology Supercomputing
Centre in cooperation with Silicon Graphics/Cray Research, supported
by HEFCE and PPARC.

\end{document}